\documentclass[dvips]{arxstspdf}
\usepackage{graphicx}
\usepackage{flushend}
\usepackage{stfloats}

\volume{25}
\issue{4}
\pubyear{2010}
\firstpage{566}
\lastpage{587}
\doi{10.1214/09-STS299A}

\begin{document}
\begin{frontmatter}

\title{A Conversation with George G. Roussas}
\runtitle{A Conversation with George G. Roussas}

\begin{aug}
\author[a]{\fnms{Debasis} \snm{Bhattacharya}\ead[label=e1]{debasis\_us@yahoo.com}} \and
\author[b]{\fnms{Francisco J.} \snm{Samaniego}\corref{}\ead[label=e2]{fjsamaniego@ucdavis.edu}}
\runauthor{D. Bhattacharya and F.~J. Samaniego}

\affiliation{Visva-Bharati University, University of California and University of California}

\address[a]{Debasis Bhattacharya is Professor of Statistics,
Visva-Bharati University, West Bengal, India, and a frequent visitor in
the Department of Statistics, University of California, Davis
\printead{e1}.}
\address[b]{Francisco J. Samaniego is Distinguished Professor,
Department of Statistics, University of California, Davis,
California 95616, USA
\printead{e2}.}
\end{aug}

\begin{abstract}
George G. Roussas was born in the city of Marmara in
central Greece, on June 29, 1933. He received a B.A. with high honors in
Mathematics from the University of Athens in 1956, and a Ph.D. in
Statistics from the University of California, Berkeley, in 1964. In
1964--1966, he served as Assistant Professor of Mathematics at the
California State University, San Jose, and he was a faculty member of
the Department of Statistics at the University of Wisconsin, Madison, in
1966--1976, starting as an Assistant Professor in 1966, becoming a
Professor in 1972. He was a Professor of Applied Mathematics and
Director of the Laboratory of Applied Mathematics at the University of
Patras, Greece, in 1972--1984. He was elected Dean of the School of
Physical and Mathematical Sciences at the University of Patras in 1978,
and Chancellor of the university in 1981. He served for about three
years as Vice President-Academic Affairs of the then new University of
Crete, Greece, in 1981--1985. In 1984, he was a Visiting Professor in the
Intercollege Division of Statistics at the University of California,
Davis, and he was appointed Professor, Associate Dean and Chair of the
Graduate Group in Statistics in the same university in 1985; he served
in the two administrative capacities in 1985--1999. He is an elected
member of the International Statistical Institute since 1974, a Fellow
of the Royal Statistical Society since 1975, a Fellow of the Institute
of Mathematical Statistics since 1983, and a Fellow of the American
Statistical Association since 1986. He served as a member of the Council
of the Hellenic Mathematical Society, and as President of the Balkan
Union of Mathematicians. He is a Distinguished Professor of Statistics
at the University of California, Davis, since 2003, the Chair of the
Advisory Board of the ``Demokritos Society of America'' (a Think Tank)
since 2007, a Fellow of the American Association for the Advancement of
Science since 2008, and a Corresponding Member of the Academy of Athens
in the field of Mathematical Statistics, elected by the membership in
the plenary session of April 17, 2008.

\end{abstract}


\end{frontmatter}

This conversation took place in George Roussas' office at the University
of California, Davis, on the 15th of May, 2009.

\section*{Early Years and Family Background}

\textbf{Debasis and Frank}: George, it's a pleasure to have this
opportunity to chat with you about your life and career. We're coming at
this conversation from different angles, one of us as a regular research
collaborator over the last ten years and the other as a long time
departmental colleague. Our common ground is that we are both long-time
friends and admirers.

Let's start at the beginning. Tell us a bit about your early days.

\textbf{George}: Let me say, first, that I'm greatly honored that you
asked me to have this conversation, and I have been very much looking
forward to it.

I was born in the city of Marmara, broadly in a family of educators.
Marmara is a small community (of maximum population of about 1350) on
the Greek mainland. It is widely thought to be within the location of
Achilles' ancient kingdom. I attended the elementary school in Marmara
and in Thessaloniki, where part of my family was. My high school
education was also divided, started in Thessaloniki and completed in
Athens. The schooling was highly structured, as was typical in Greece,
and very rigorous. The environment in Marmara was idyllic, and I still
have very fond memories of it.

\begin{figure}

\includegraphics{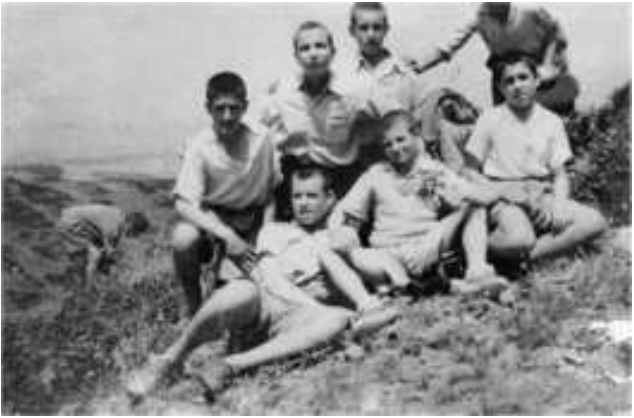}

\caption{George Roussas (in the middle of front row) in the 5th or 6th grade in Thessaloniki,
1944--1945.}
\end{figure}

\textbf{Debasis and Frank}: What can you tell us about your family
background?

My paternal grandparents had three sons (my father Gregory and my uncles
Hercules and Constantine). These two uncles obtained university degrees,
but my father was business oriented. In the early 1920's he left Marmara
and went to Thessaloniki, where he entered in the dairy business. He had
considerable success in this, holding a prominent place in the
distribution business in Thessaloniki for almost two decades. Indeed, he
was a self-made millionaire! My parents (Gregory and Maria) had four
children, daughters Aggeliki, Demetra, and Stella, and myself. Our
family moved to Thessaloniki, and later to Athens, ``in stages,'' as the
locations where my father's varied business interests were centered
changed over the years. Of course, a civil conflict in the country had
its influence on this as well.

My father was a product of the classical European liberalism, which in
the 1920's and early 1930's was well represented in Greece by a
remarkable statesman, Eleftherios Venizelos. My father was an active
member of the liberal party and a staunch supporter of Venizelos. He
retained this position throughout the late 1930's, when Greece was run
by a politician turned dictator, and during World War II, and later,
when Greece was savaged by a civil war. His political liberalism and
outspokenness cost him dearly. During the German occupation of the
country, he was sent to prison (released as a political prisoner at the
end of the war, upon the liberation of the country). During part of the
civil war, he was exiled to a remote deserted island by the governing
party. The fact that my father saw some glitter of light in the
repressive soviet system, such as the availability of abundant
opportunities for well-qualified students to pursue their educational
goals, did not sit well with the party in power at the time. From a
financial viewpoint, he believed strongly in currency rather than in
property. Consequently, his hard-earned ``fortune'' became worthless
with the nullification of the Greek currency during World War II, and a
millionaire became virtually penniless! It was in this kind of
environment in which I grew up. This environment shaped my determination
to distance myself from any business per se and to pursue education to
its highest level possible.

\section*{Becoming a Mathematician while Searching for Ithaca}

\textbf{Debasis and Frank}: What were your main interests when you
entered college? Did you have strong feelings about what you wanted to
specialize in?

\textbf{George}: In high school, I developed a strong affinity to the
humanities and social sciences, with margi\-nal interest in mathematics
and physical sciences. Soon, I~discovered that any weaknesses in mathematical and
physical sciences would deprive me of many options in later years. So, I
decided to intensify my efforts, and graduated with a strong record in
all my subjects. This standing put me on a solid position to compete for
a position in the air force academy; it was my youthful dream to become
an air force officer. But that dream would never come to fruition. In
addition to succeeding in a competitive examination, I would also have
to have the written consent of both of my parents. I~thought I could
talk my father into it, but my mother was adamantly opposed to the idea.
Instead, I was advised by my uncle Hercules (the dean of the
classicists, as he was often referred to) to take the entrance
examination in the department of mathematics at the University of
Athens. Reluctantly, I took his advice, but I~never checked the results
of the entrance examination. My fixation was still with the air force,
and at this time, I targeted the aeronautical engineering school of the
air force. However, there seemed to be a problem here. Namely, those
competing for a position were more than 300, and the positions available
were 6--8! In view of these imposing odds, my parents did not attempt to
dissuade me from preparing for such a competition, and actually taking
the examination. Why should they? It was, clearly, a hopeless effort!
For about a year, I exhaustively studied math and science. When the
examination time arrived, I was an enthusiastic and determined
participant. In military schools, the exams were taken serially, and
only the successful participants in one subject were allowed to continue
with the next subject. In this manner, I reached the last examination in
chemistry, which was taken by less than a couple dozen people. I later
learned that I had earned the top overall score on the examination.

And it was here when the drama began. Succeeding in the examinations was
extremely tough, but that was only part of the admissions routine. The
candidates for all military schools, and, in particular, for such an
elite institution as the air force aeronautical engineering school, also
had to be certified on their political beliefs, on the basis of several
degrees of family connections. It was here where the sorry political
past of my father entered the picture. As became known later, the
disqualifying certificate arrived at the examination committee's
headquarters right after the examination papers in chemistry were
corrected. It was the duty of the committee to flunk me, no explanations
provided. The chairman of the committee, an air force colonel-engineer,
took it upon himself not to post the results. My uncle Hercules, who was
highly regarded and had many influential acquaintances, tried vigorously
to obtain an exception for me, unfortunately, to no avail. He was given
to understand that there would be dire political consequences if the
effort to gain my admission to the elite school of aeronautical
engineering of the air force succeeded. So, I was officially certified
to be a$\ldots$communist (!), and I was abruptly denied the realization
of my dream.

\textbf{Debasis and Frank}: Disappointing and demoralizing! What
happened next?

\textbf{George}: For more than a year, my odyssey in search of my Ithaca
went on, without much satisfaction. After quite some time, disappointed
and shaken, I decided to visit the department of mathematics of the
University of Athens, just to inquire about the previous year's entrance
examination. I was told that I was successful, but since I did not
enroll, I lost the right of enrollment. Fortunately, there were a couple
of openings in the following year's class, and one was allocated to me.
Apparently, my manifest destiny was to become a mathematician rather
than an air force officer-engineer! I~graduated from the University of
Athens in four years with high honors. During the last two years, I also
served as a teaching assistant to professor D.~A. Kappos, who was a
student of Constantine Carath\'{e}odory and held the chair of
mathematical analysis. I was fortunate to take many of my courses from
him. While my studies in mathematics were quite broad, I had not yet
been introduced to probability or statistics.

\begin{figure*}

\includegraphics{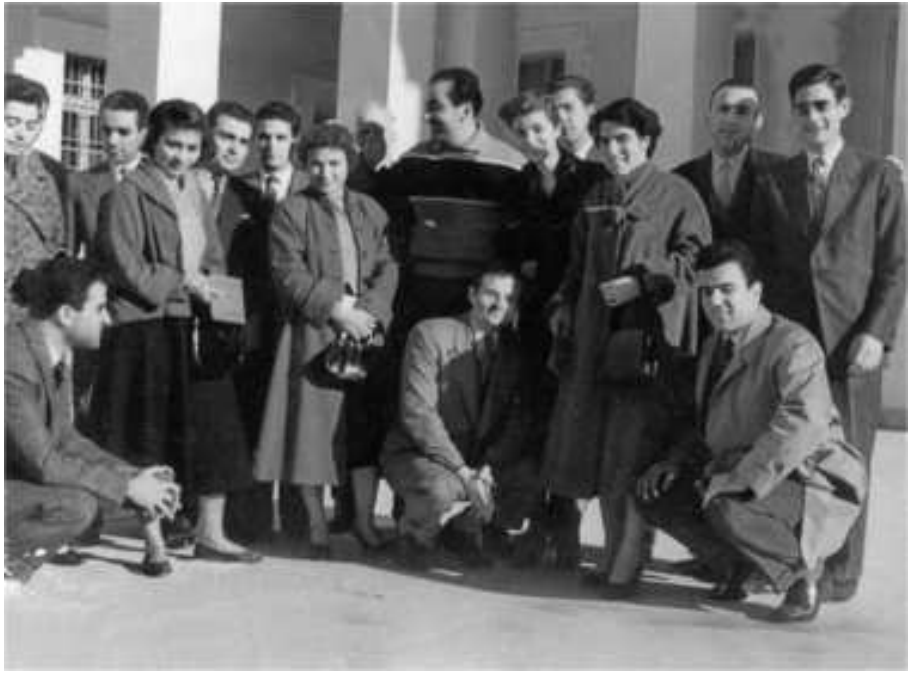}

\caption{A (nonabelian!) group of the mathematics graduating
class at the University of Athens, fall 1956. (George Roussas, in the
middle, kneeling).}
\end{figure*}

\section*{Choosing Statistics---The University of California, Berkeley Years}

\textbf{Debasis and Frank}: That's a fascinating story, Geor\-ge. It's
interesting how sometimes bad things seem to happen for a reason.
Certainly your ultimate career path is a good example. It's intriguing
that you decided to pursue graduate work in a field that you had yet to
be formally exposed to. How did  that come about?

\textbf{George}: It was my determination to pursue graduate work abroad.
The decision to go for statistics---despite the lack of any relevant
background---was due to a liking I took in probability (by attending an
occasional seminar, and also by studying on my own), but primarily it
was due to the advice of Professor Kappos. His own expertise was in
measure theory and probability in abstract structures. The absence of
statistics from the curriculum was an additional reason. Since studying
abroad was well beyond my family's financial means, another financial
source would have to be located. Fortunately, the Greek government did
provide some relevant fellowships, based on a series of written
examinations and service in the armed forces. So, I was inducted into
the army, where I served for two years as a private (an unusually low
rank for a young man with my background), largely because I was still
plagued by my unfortunate experience with the air force. But my low rank
army service did me some good, as I spent almost the entire period close
to home, and I had the possibility to pursue my study of the English
language. Soon after my discharge from the army, I participated in an
examination for the selection of fellows to study applied mathematics
(which included probability and statistics) abroad. Professor Kappos
insisted that, once the decision to study statistics was made, the place
to go would be the University of California in Berkeley.

\begin{figure}[b]

\includegraphics{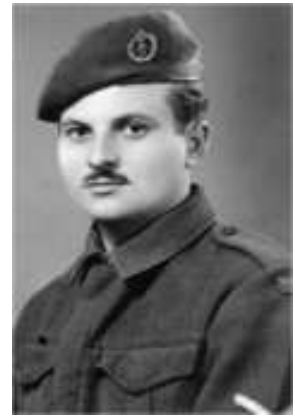}

\caption{George Roussas when serving his military service in the Greek
Army, 1957--1959.}
\end{figure}

\textbf{Debasis and Frank}: You began your graduate studies in the
states in 1960. What were your first impressions of Berkeley?

\textbf{George}: I traveled from Athens to Berkeley by way of London and
New York. My first impression of New York was awful, but California and,
in particular, Berkeley, was another matter. The climate is about the
same as that of southern Greece, the city of Berkeley is charming, and
the university campus is of exceptional beauty. Nothing, of course, is
needed to be said about the academic standing of the entire university,
and of the department of statistics, in particular. David Blackwell was
the chair of the department, and Lucien Le Cam was the graduate advisor.

\begin{figure}

\includegraphics{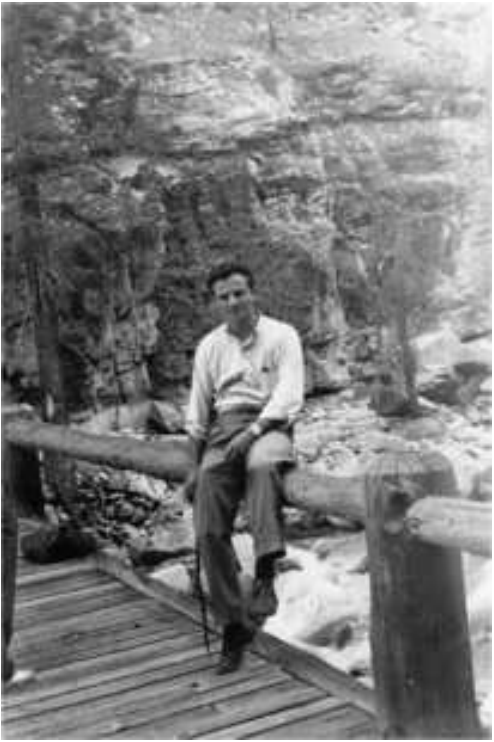}

\caption{George Roussas in the Yosemite National Park in his first summer
in the UC Berkeley, 1960.}
\end{figure}

\textbf{Debasis and Frank}: Tell us a bit about your graduate studies
and, in particular, about the faculty members who had a strong impact on
you.

\textbf{George}: The first graduate level course in probability and
statistics I took from Thomas Ferguson, during the summer session. Early
on, I took courses in measure theory, functional analysis and topology
from the department of mathematics. The hypotheses testing and point
estimation courses I took from E.~L. Lehmann, and decision theory form Le
Cam. I took courses in measure-theoretic probability, second order
processes and sufficiency form Edward Barankin. I took a one-year course
in probability from Lo\`{e}ve, and a course in Markov chains from David
Freedman. From Blackwell, I took a course on coding theory and another
one on programming. Also, I took a course in Markov chains from K.~L.
Chung during a summer, and another course in empirical Bayes methods
from Herbert Robbins when he was a visitor in Berkeley. It was almost a
criminal omission that I did not take the ANOVA course from Henry
Scheff\'{e}, and at least one of the courses taught by Jerzy Neyman. I
did, however, study the Scheff\'{e} book thoroughly and, later, taught
out of it.

The faculty of the department of statistics in the UC Berkeley was an
almost$\ldots$suffocating constellation of stars! I had immense respect
and admiration for each and every faculty of the department. Neyman---founder
of the statistical laboratory and of the department of
statistics---was an imposing figure in the department. He was very kind
to me, and more than once mentioned to me his experience during a brief
visit in Greece as an international observer. Barankin, in addition to
being an outstanding mathematical statistician and probabilist, was also
well versed in philosophy and in the classics. It was not unusual for
him and me to talk about Plato, Aristotle and Sophocles. I learned
asymptotic theory primarily from Le Cam. His seminal work on contiguity
of sequences of probability measure and its statistical implications
were the key for my entrance into the field of large sample theory. As
is well known in the statistical community, Le Cam was deeply
knowledgeable in a broad area of mathematical sciences, and exceedingly
helpful to all those who sought his advise. Le Cam's vast knowledge
often created communication problems between himself and a student.
However, those who persisted would manage eventually to chip away bits
of his wisdom.

David Blackwell was my great discovery at UC Berkeley. It is not a
secret that  UC Berkeley was the repository of great scientists. So,
in this context, it would not come as a surprise that Blackwell belongs
in that exclusive club. What is rather rare, however, is for a great
scientist to be endowed with exceptional human qualities. That is,
indeed, the case, which puts Blackwell in a class of his own. He is
endowed with a refined, friendly and appealing personality, and he
treats people in ways that build their self confidence and inspires
relationships based on mutual respect. He's been a wonderful role model
for me and many others.

As one would expect, the time in the UC Berkeley was academically
challenging, but overall pleasant, and certainly extremely constructive;
it provided unmatched academic training.

\begin{figure}

\includegraphics{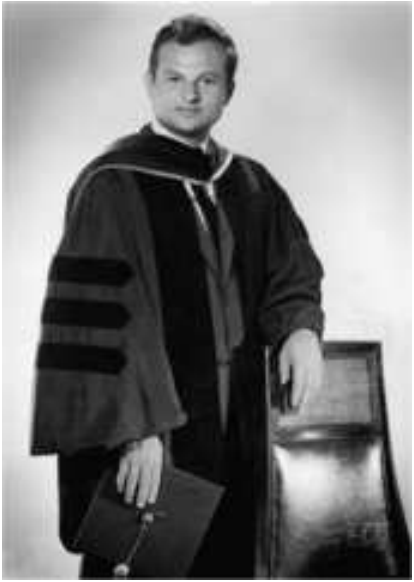}

\caption{George Roussas upon his graduation from the UC Berkeley, 1964.}
\end{figure}

\begin{figure*}

\includegraphics{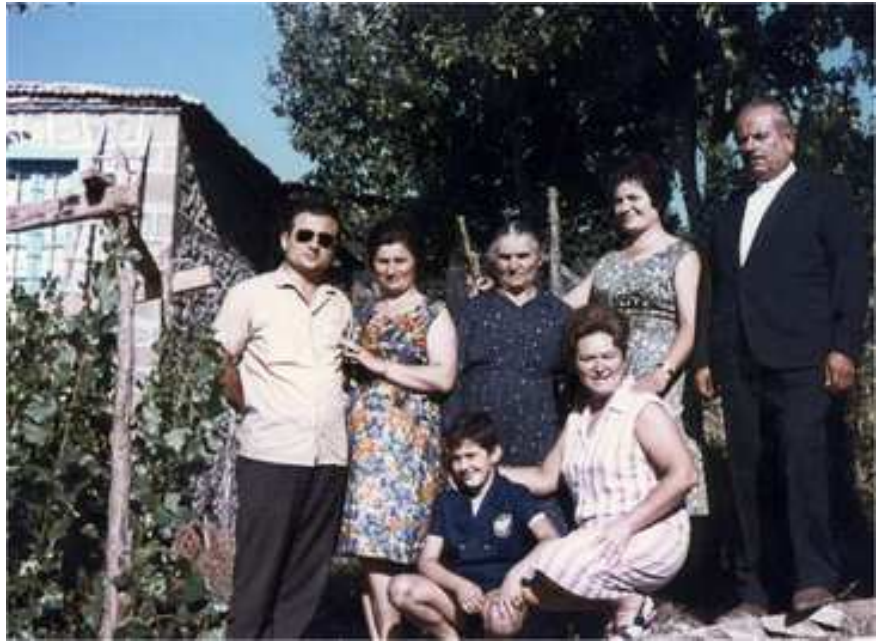}

\caption{George Roussas' family in Marmara, in the summer of 1966. From
left to right: George, sister Demetra, mother, sister Stella, father,
sister Aggeliki (kneeling), and nephew John.}
\end{figure*}

\section*{The University of Wisconsin, Madison, Experience}

\textbf{Debasis and Frank}: After taking a temporary position in 1964
(while considering a possible return to Greece), you joined the
Statistics faculty at the University of Wisconsin, Madison. George Box
was then in the early stages of organizing the statistics department
there. What were the highlights of your time in Madison?

\textbf{George}: In the fall of 1965, I was invited to interview at UW
Madison. At the end of my interview, Irwin Guttman, then the acting
chair of Statistics there, made me an unofficial offer, and I accepted
it on the spot. I had already fallen in love with Madison, both because
of its physical beauty and because of the superb academic climate there.
I did not allow myself the time for the usual bargaining to improve upon
the rather low academic salaries offered by the UW at the time!

So, I joined the department of statistics of the UW Madison, in the fall
of 1966, as an assistant professor. At the same time, another four
assistant professors were hired: Asit Basu, Richard A. Johnson, Gouri
Bhattacharyya and James Bondar. Existing faculty members, in addition to
Box and Guttman, were Norman Draper, John Gurland, Bernard Harris,
William Hunter, Jerome Klotz, George Tiao, Donald Watts and Sam Wu, as I
recall.

George Box had created in the department an academically demanding and
rigorous climate, but at the same time, comfortable, nonoppressive, and
enjoyable. Those who did their work well were recognized and rewarded. I
was promoted to associate professor (with tenure) in 1968 and to full
professor in 1972. As all other faculty members, I used to teach two
courses per semester, one graduate-level course and one undergraduate
course. The undergraduate course was alternated between an upper
division probability and mathematical statistics course, and a
pre-calculus statistics course. The latter choice was often made,
because that was where the interesting students were! We recruited some
very good ones into the statistics major!

\begin{figure*}

\includegraphics{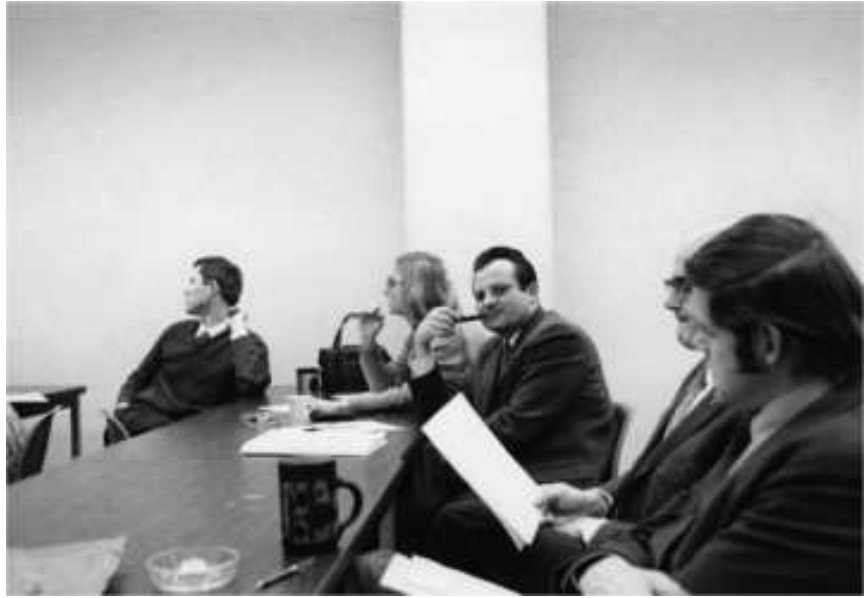}

\caption{A faculty meeting in the Department of Statistics of the
University of Wisconsin, Madison, sometime between 1968 and 1970. From
left to right (part of the faculty only): Jerome Klotz, Grace Wahba,
George Roussas, John Gurland and John Van Ryzin.}
\end{figure*}

\textbf{Frank}: I understand that at least one of them was recruited
into marriage! (Laughs.)

\textbf{George}: You are right about that! It was in one of my
pre-calculus statistics classes that I met Mary Louise Stewart, who was
destined to become my wife. She was a Ph.D. candidate in food management
with a minor in statistics. This was in the fall of 1969. During the
spring semester of 1970, I was on sabbatical leave, which I spent in the
famous mathematics institute of the University of Aarhus in Denmark as a
guest of Barndorff-Nielsen. It was also there that I wrote the draft of
my book on contiguity and where I met the great K. It\={o} and attended
his ergodic theory seminar.

Sometime early in the fall of 1970, after my return to Madison, I
contacted Mary, and she responded positively. We started dating
regularly, and were engaged in the summer of 1971. I took Mary to Greece
to meet my parents, sisters and close relatives in the summer of 1971,
and upon our return to Madison, we had our civil wedding ceremony on
September 11, 1971.

\begin{figure}[b]

\includegraphics{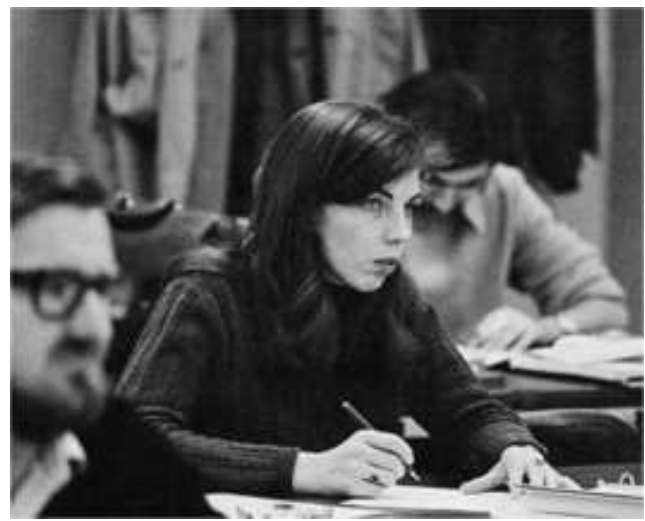}

\caption{Mary Roussas in George Box's class on Time Series Analysis as a
graduate student at the UW Madison when she was still Mary Stewart,
1970.}
\end{figure}

\begin{figure}[b]

\includegraphics{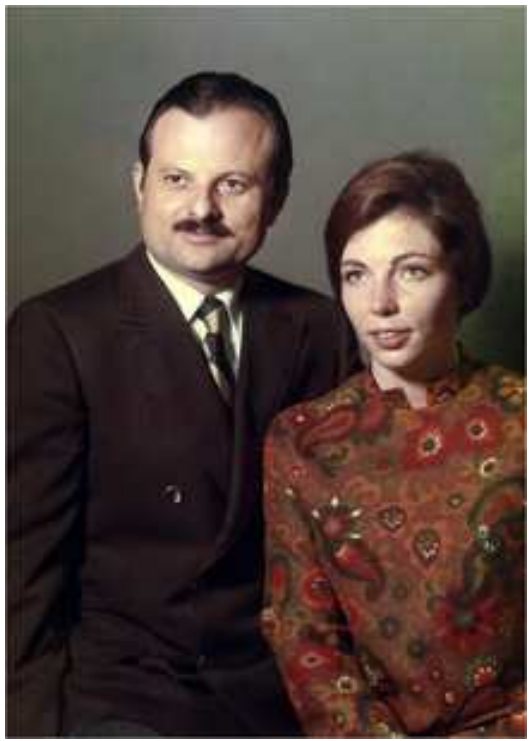}

\caption{George and Mary Roussas newly married in Madison, Wisconsin,
1971.}
\end{figure}

\section*{The Pending Issue of Returning to Greece---Traveling Between Madison
and Patras}

\textbf{Debasis and Frank}: Things went quite wonderfully for you in
Madison, both personally and professionally. We know, however, that you
were faced with a difficult choice in the early 1970s---whether to
remain at Madison or return to your native Greece. What were the main
issues you were dealing with at that time?

\textbf{George}: Madison was great for us in so many ways. However,
there was a recurring issue that cau\-sed quite a bit of discomfort. I
began to receive repeated notifications from the Greek government about
my contractual ``obligation'' to return to Greece and my need to
discharge this obligation. At the time, Greece was under military rule,
and that made my return there problematic on many counts. From a purely
practical viewpoint, I was highly content with my life and career, and I
had no desire to leave Madison. Further, Mary and I were already
planning to start a family. Philosophically, I was strongly opposed to
serving under a military regime. Finally, a return to Greece seemed
unsafe to me, as I had been active in opposing the military regime. On
the other hand, I had no wish of being deprived of my Greek citizenship,
as was being threatened. Incidentally, I became an American citizen on
May 28, 1971, and ever since, I have been grateful to the American
people for the privilege of citizenship bestowed upon me.

\textbf{Debasis and Frank}: So, how did you resolve this vexing
conflict?

\textbf{George}: I decided to respond to demands made with a proposal
that seemed like it had virtually no chance of being accepted. I
indicated that I could consider a return to Greece only if I was offered
an academic position there comparable to the one I was holding in
the States. Since there were quite a limited number of professorships in
Greek universities at that time, and occupying a full professorship was
only for the well connected, the possibility seemed, to say the least,
remote. So, on this account, I felt fairly safe. Unfortunately (for me),
the government came up with an open chair in Applied Mathematics (which
included probability, statistics, numerical analysis and a few other
subject matters) in the new and promising technologically oriented
University of Patras (UP) (situated about 150 miles west of Athens), and
insisted that I submit a candidacy. Still feeling safe for the reasons I
cited above, I submitted an application, and lo and behold, I was
elected (to a full professorship). However, the Minister of Education
refused to ratify the election and ordered for the chair to be declared
open again. At this time, electors (all full professors of the School of
Physical and Mathematical Sciences) pleaded with me not to object to the
resubmission of an application on my behalf. The process of election
commenced anew, and I was elected again! This time, the Ministry of
Education kept the outcome of the election in its drawers---neither
rejecting nor ratifying it---for a few months, until a new Minister, a
civilian, came in to replace the previous person, who was a military
officer. This fellow was a chemical engineer with a Ph.D. degree from
McGill University, who had worked for the Shell Oil Company in the
States over many years. As soon as he was informed about the long
pending ratification of my election, he approved the election
immediately. That was the right thing for him to do, but it did not
serve my purposes well. I was strongly urged to go to Athens to take the
oath of the office, and perhaps be given leave of absence for a limited
period of time. The compromise reached was to take the oath of office in
the Consulate General of Greece in Chicago, and report for duty in early
1972.

Under these circumstances, Mary selflessly abandoned her studies
temporarily (she had already taken her Master's degree, and was well on
her way toward fulfilling all requirements for the Ph.D. degree), took a
crash course in the Greek language, and started preparing herself for
the forthcoming adventure. The colleagues in the department attempted to
dissuade me from going to Greece, and insisted that I retain my
appointment at Madison while taking a leave of absence of indeterminate
duration. I have always appreciated this gracious gesture.

\textbf{Debasis and Frank}: So, this is when your triumphant, if
somewhat reluctant, return to Greece began.

\textbf{George}: In a manner of speaking, yes. In February 1972, Mary
and I departed for Patras. Now, the city of Patras and the university
campus are built in a beautiful location, on slopes overlooking a bay,
with the western part of Greece opposite it. However, being acclimated
in a new community (and, for Mary, a foreign community at that) did pose
considerable problems. In the university itself, I was received well by
some, and not favorably by others. In dealing with the authorities, both
in Patras and Athens, my strong point was that I had a safe escape route
and, therefore, I had no problems in behaving in my natural way. I
started teaching immediately courses in probability and statistics,
organizing the Laboratory of Applied Mathematics, recruiting TA's and
personnel for the lab, mentoring students with an interest in
probability and statistics, attending hours-long and stormy faculty
meetings, etc. Mary made a valiant effort to adjust to the local
conditions, and tried hard to improve her Greek vocabulary. I am happy
to say, though, that she was treated exceptionally nicely by all
involved.

In the fall of 1972, we returned to Madison, and in the winter back to
Patras. In the fall of 1973, we returned again to Madison. Mary also
gave birth to our first son, Gregory, that October 18. Unfortunately,
soon thereafter, I had a rather serious operation (removal of slip
discs) in the University Hospital, and I could offer little to the
department at that time. By the end of the year, we returned to Patras,
where I completed my recuperation.

\textbf{Debasis and Frank}: Traveling between the two pla\-ces must have
been quite cumbersome!

\textbf{George}: Yes, the moving back-and-forth between Madison and
Patras eventually necessitated for us to essentially retain two
households. It was financially challenging and physically tiring. At the
beginning, it was Mary and me (and Mary's three cats from her student
days!), and now it was Mary, me and Gregory (in addition to the three
cats!). It was clear that a decision was due soon as to where we were to
affix our affiliation. At that time, everything pointed toward Madison,
as my academic experience in Patras had been a disappointment to me. In
addition to all my painful efforts to organize and staff a new unit, I
repeatedly encountered what I considered to be harassment from the
Minister of Education (who seemed intent on imposing political
considerations into university affairs). In time, we were able to
establish a tentative truce, allowing me to proceed with academic
matters in ways that American academics consider natural and perhaps
sometimes take for granted.

It was about this time that a decision about returning to Madison
permanently was due, when all of a sudden the military regime collapsed
(in July 1974), and a civilian government took over. The people, by and
large, were elated with the change. A military regime is not a normal
and natural regime for free people, in particular, for the country where
the concept of democracy was invented and first practiced. Nevertheless,
it appears that every regime has its excesses. Even the new civilian
regime was eventually credited with its share of excesses, in
particular, in the academic world. In any case, despite some
shortcomings, the new civilian regime was responsible for establishing a
semblance of normalcy in Greece and for opening up some new horizons.
This promising outlook led me to decide to remain in Greece and to
resign my appointment at the University of Wisconsin, Madison. I did
this in the fall of 1976, thus culminating four and a half years of a
joint appointment between the two institutions. It also, by no means
painlessly, terminated a ten-year association with the great university
and beautiful city in which I had met the love of my life, flourished
personally and professionally, and spent the most enjoyable years of my
academic career.

\section*{The Experience in Greek Universities}

\textbf{Debasis and Frank}: The next chapter in your professional life
was spent as a faculty member and administrator within the Greek
university system. Tell us about those years.

\textbf{George}: Yes, I was now fully identified with the University of
Patras. I felt that it was incumbent upon me to do all I could for the
benefit of the institution, while also looking after my own scientific
survival. I~had worked hard seeking out the best available candidates
(of Greek descent, as required) whenever a faculty position became
available. I expanded this effort to the entire spectra of biological,
natural and physical sciences. These kinds of activities were not
universally appreciated, but I was nonetheless narrowly elected as the
Dean of the School of Physical and Mathematical Sciences (by the full
professors of the school). Around this same time, Mary became pregnant
with our second son, John. He was born in Bloomington, Indiana, on
August 10, 1977, while I was spending the summer as a research professor
at Indiana University.

\textbf{Frank}: Is that when you were offered a starring role in the
bicycling classic film ``Breaking Away''?

\textbf{George}: No, that came later! (Laughs.) This time, I just went
to visit and work with Madan Puri, an old friend of mine since our UC
Berkeley days, and my former student from Greece, Michael Akritas, now
an outstanding senior statistician, as we all know.

\textbf{Debasis and Frank}: Following this leave, you returned to Patras
to take on the deanship with renewed energy?

\textbf{George}: Exactly! I did not feel bound by academic traditions
that didn't seem to work. My main guides were common sense and my
experience with US universities. One of my early ``accomplishments'' was
to reform the manner in which faculty meetings were run. Regular
school-wide faculty meetings were always held in a large room with the
faculty seated, according to academic seniority, around a huge oval
table. Meetings were seen as both business and social affairs. The
agenda was typically unrealistically long, so that meetings dragged on
and on for hours, often without any truly useful work being done. I
introduced a new system in which the agenda was prioritized, placing
first the items on which faculty input was essential. I made a strong
effort to exclude items which were politically driven, and to set aside
strictly administrative issues which could be handled without taking up
the faculty's time. While exerting control of the agenda, I resolved to
be firm but also fair and impartial.

\textbf{Debasis and Frank}: By the way, reconnecting with your family
during this period must have been a special pleasure.

\textbf{George}: Most certainly so! My parents especially enjoyed seeing
our children on a regular basis.

\begin{figure}

\includegraphics{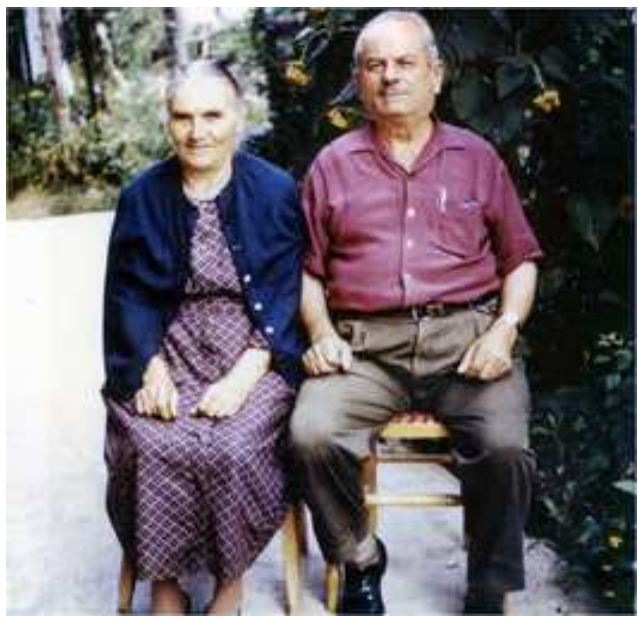}

\caption{George Roussas' parents in the late 1970's.}
\end{figure}

\textbf{Debasis and Frank}: How was your approach to the deanship taken
by the faculty in Patras?

\textbf{George}: While my methods were considered new and different,
most of my colleagues were pleased with my performance and some of them
urged me to stand for election for the office of the chancellor of the
university. In those days, the chancellor was elected by the totality of
the full professors of the university; this was the old continental
European system. I was elected by a comfortable margin. I served as a
chancellor-elect for a year, and took over the chancellorship the year
after.

\begin{figure*}

\includegraphics{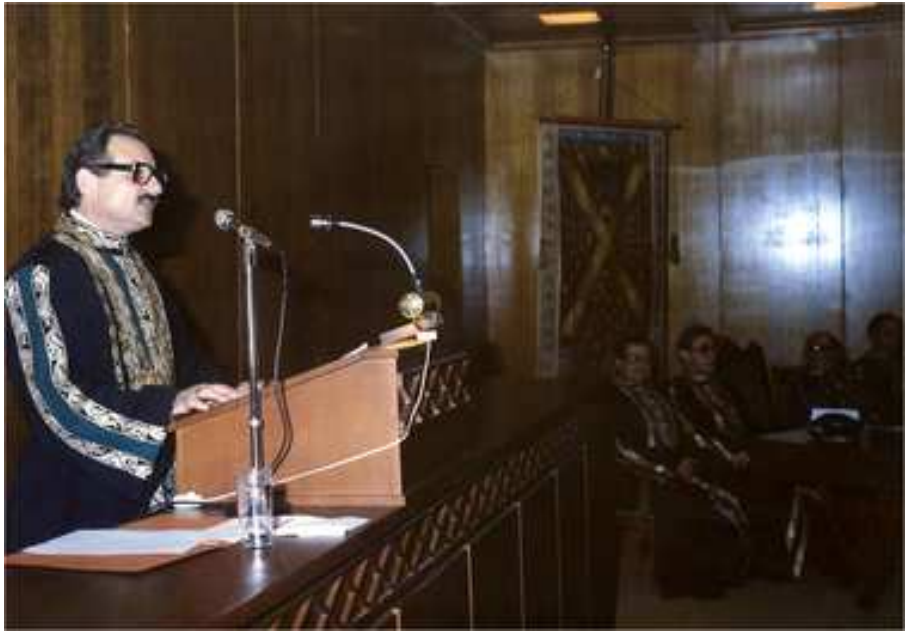}

\caption{George Roussas delivering a speech at the University of Patras
at his inauguration as the Chancellor of the university, 1982.}
\end{figure*}

\begin{figure*}

\includegraphics{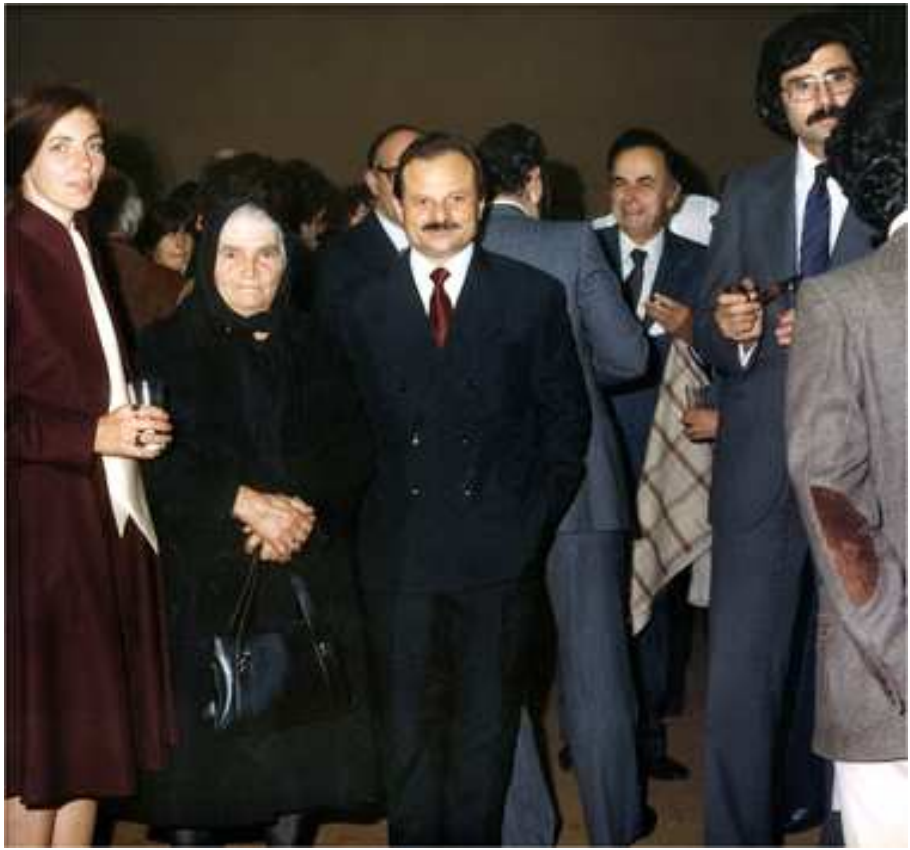}

\caption{At the reception right after George Roussas' speech. From left
to right: Mary Roussas, George Roussas' mother and George Roussas. (His
father had passed away).}
\end{figure*}

\textbf{Debasis and Frank}: An important political change began in
Greece in 1982 when A. Papandreou formed a new government. What did you
know about him at the time?

\textbf{George}: Papandreou came from a political family (his father was
the leader of a political party, and had served both as a minister and
prime minister in the past). He left Greece right after high school,
studied economics in Harvard, and served on the faculty of several US
universities, most notably UC Berkeley, where he was also the chair of
the department of economics for some time. It was there where I came to
know him. He was a noted economist, clearly a political leader with
highly respected credentials within the European Union (EU). The new
government was welcomed by a substantial majority of people as a turning
point in Greek politics, and justifiably so. A modern and knowledgeable
economist at the helm of the government would surely put the vast
amounts of resources flowing form the EU to good use, developing and
modernizing the Greek economy. It was generally expected that a man of
his background would also revitalize Greek education at all levels,
helping to recruit a substantial number of Greek academics from home and
abroad, thereby infusing Greek universities with new blood and highly
qualified scientists.

\textbf{Debasis and Frank}: Around this same time, you yourself made a
change within the Greek University system. Tell us about that.

\textbf{George}: Actually, this was not a change of my base, it was the
undertaking of temporary additional duties. Undersecretary of Education
George Lianis offered me the position of the vice president for academic
affairs of the then new University of Crete. The appointment would allow
me to remain in Patras as chancellor, but it would require weekly
meetings, in either Athens or Crete. My primary new responsibility was
to oversee and chair the elections of faculty members in the Departments
of Mathematics, Physics, Chemistry, Biology and Computer Science. I felt
I could accommodate these duties in my portfolio without disturbing my
family, now grown to include Mary and three sons, with the addition of
George-Alexander on December 12, 1980. I launched into my new
responsibilities with gusto, and before long, the University of Crete
was staffed by scientists of considerable international reputation.
There were some challenges to be faced, but I'm saving the details for a
mystery novel I will compose in the near future! (Laughs.)

\begin{figure*}[b]

\includegraphics{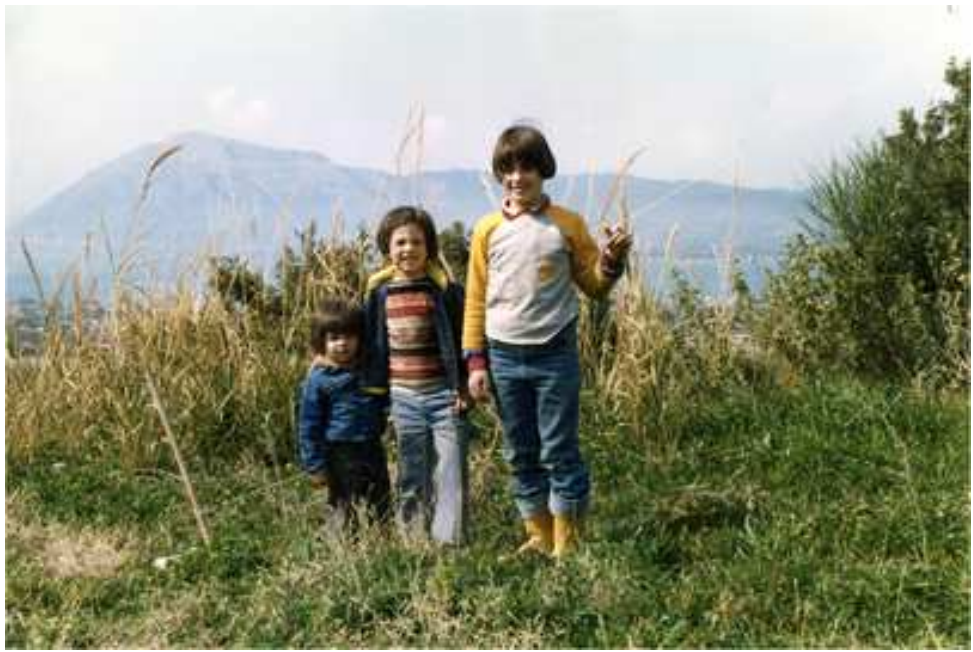}

\caption{The Roussas boys. From right to left: Gregory, John and
George-Alexander, in Greece, 1983.}
\end{figure*}

Something about the Greek system of administration might interest you,
as things are quite different in the States. Neither the deanship nor
the chancellorship I held provided any additional payment or stipend
beyond the professorial salary. As a chancellor, I had at my disposal a
car and chauffeur for university business around the clock, and that was
all. For my service in the University of Crete, I was simply paid travel
expenses and a nominal per diem. In Greece, academic administration is
viewed as within the normal scope of a professor's activities.

\textbf{Debasis and Frank}: What seemed like a very promising beginning
at Patras and at Crete was, unfortunately, destined to run into
insurmountable obstacles. What were the root causes of the change in the
academic climate in Greece?

\textbf{George}: Perhaps not unexpectedly, politics trumps most other
forces in our society. While the conditions were ideal to usher Greece
into a new era of achievement and prosperity, it soon became clear that
this was not going to happen. It became apparent that Papandreou had no
intention of being the architect of such a feat. What he did instead was
to continue preaching and practicing his pre-election populism, even
after he was in power. He squandered the national wealth and the
resources provided by the EU for partisan purposes and other unworthy
causes; consequential productive investing was nowhere to be seen.

The education system of the country, and, in particular, the higher
education, was in dire need of reorganization. Papandreou's
``reorganization'' essentially abolished the administrative structure in
the primary and secondary schools, and virtually dismantled the
universities. The existing small number of university professors was
marginalized with the flooding of the ranks by new appointees, and the
universities were turned into unlikely arenas of competition of the
political parties. By political collaboration of teachers and students,
the resulting majority was then in a position to elect the university
authorities at all levels (departmental chairs, deans, vice chancellors
and chancellors). Needless to say, the result was predictable chaos and
a dramatic lowering of academic standards. It has been more than a
quarter of a century since these measures were put into effect, and the
results are everywhere to be seen. Furthermore, there is no hope for
deliverance form this evil anytime soon; the genie is out of the bottle,
and it is not easy (or even possible) to confine it in there again!

\section*{The Turning Point---Returning to the States}

\textbf{Debasis and Frank}: That is indeed a tragedy. It's clear that
you harbor both sadness and anger about it; sadness for your native land
and anger about the way things were changed for the worse. As you
completed your term as chancellor of the University of Patras, you could
see the handwriting on the wall. That's about the time that you took a
sabbatical leave at the University of California, Davis, is it not?

\textbf{George}: That's exactly what happened. Although our next move
was not yet clear, Mary and I decided that a year of sabbatical leave,
spent outside the country, would be a welcome and much needed change,
and help us work out a plan for the future. That future could have been
a suitable position in the EU. Nevertheless, we decided to spend the
year in the States, and that is how I found myself at the UC Davis in
the capacity of visiting professor, starting in the summer of 1984. P.~K.
Bhattacharya's work on nonparametric statistics was one of the reasons
that I was drawn to UC Davis. Statistics at UC Davis at that time was
organized as an Intercollege Division of Statistics headed by an
associate dean. Professor Robert Shumway was the acting associate dean,
and he was prompt and most accommodating in his response to my request
about visiting the UC Davis for the year.

As you well remember, Frank, the UC Davis statistics unit was formed in
1979 in the usual manner, that is to say, by grouping together
statisticians affiliated with other departments, such as mathematics,
epidemiology, etc. It was a solid group of fair size, and its first
associate dean was Julius Blum, a noted probabilist. Other members of
the unit in the early 1980's were P.~K. Bhattacharya, Alan Fenech, Wesley
Johnson, Y.~P. (Ed) Mack, Norman Matloff, Frank Samaniego, Robert
Shumway, Jessica Utts, Alvin Wiggins and Neil Willits. Jane-Ling Wang
came aboard the same year with me in 1984. The idea behind this mode of
organization of the unit, that is, as an Intercollege division of
statistics rather than a department of statistics, was to gather
together all statistical activities under one roof, rather than having
them spread over the campus. In the UC Davis there is also a rather
novel idea at work, that of a Graduate Group, which brings together
faculty with common research interests serving in various units on
campus. Actually, it is the graduate group which controls the graduate
curriculum and supervises graduate degrees. So, there also was a
graduate group in statistics, and the associate dean of the intercollege
division of statistics was, ex officio, the chair of the graduate group.
Blum passed away unexpectedly in his third year as associate dean of the
unit, and Professors Bhattacharya and Shumway reluctantly served in
succession in an acting capacity, while an active search was launched
for a permanent appointee as associate dean. As I recall, Frank, at that
time, you were serving in a campuswide administrative capacity as the
Assistant Vice Chancellor for Academic Affairs.

\section*{The University of California, Davis Years}

\textbf{Frank}: It seems that the stars were aligned that year, as Davis
was searching for a new head of its Statistics unit and you were
seriously looking for a new position and new challenges. I clearly
recall that you were the unanimous choice of the Statistics faculty in
our search in 1984--1985. You joined the Intercollege Division of
Statistics in July, 1985, as Professor, Associate Dean and Chair of the
broadly based Graduate Group in Statistics. You served as the head of
our unit for 14 years without taking even one quarter of sabbatical
leave. Your service to the unit was both visionary and very effective.
From your perspective, what were the highlights of this period?

\begin{figure}

\includegraphics{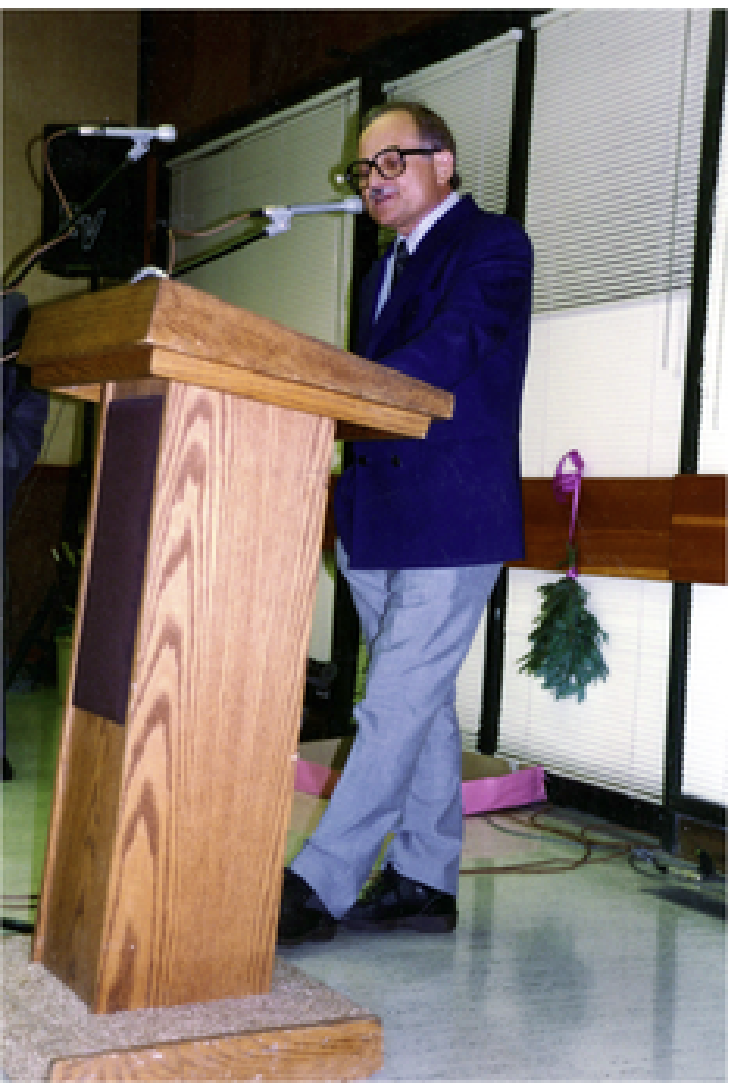}

\caption{George Roussas giving a seminar after his appointment as
Professor, Associate Dean and Chair of the Graduate Group in Statistics
at the UC Davis, 1985.}
\end{figure}

\renewcommand{\thefigure}{14.A}

\begin{figure*}[b]

\includegraphics{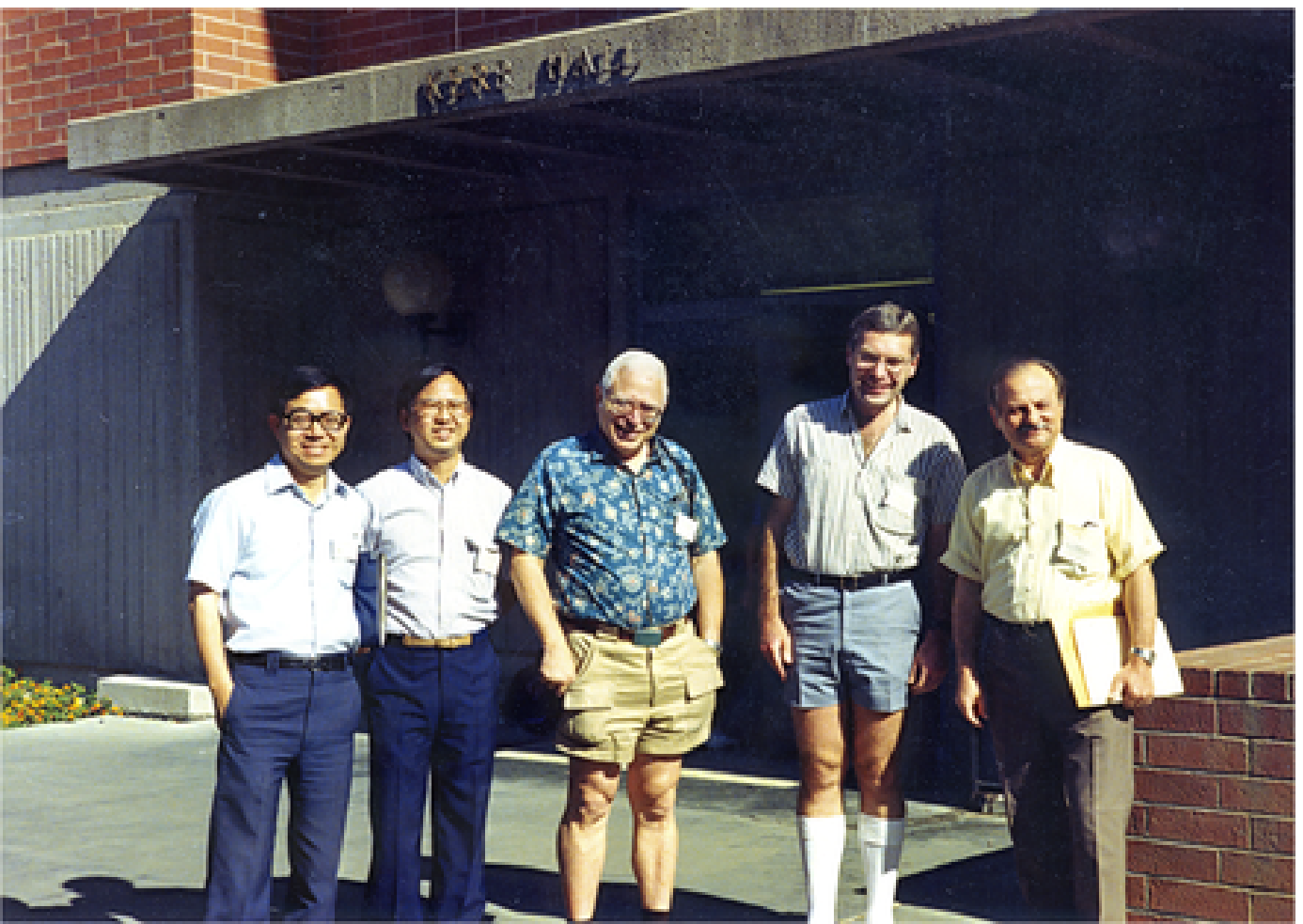}

\caption{In front of Kerr Hall at UC-Davis, in 1986. From left to
right: Keh-Shin Lii, UC-Riverside; Y.~P. (Ed) Mack, UC-Davis, Murray
Rosenblatt, UC-San Diego; Peter Hall, Australian National University;
and George Roussas.}
\end{figure*}

\textbf{George}: Thank you, Frank, for your generous description of
those years. Upon shouldering the leadership responsibilities in 1985,
the faculty, in conjunction with the university administration, designed
a strategic plan to expand the unit up to the point of achieving a
critical mass, and turn it from a solid unit to a unit of national and
international standing. We proceeded with the implementation of the plan
by hiring a number of bright new faculty members, including Prabir
Burman, Chris Drake, Hans-Georg Mueller, Wolfgang Polonik and Chih-Ling
Tsai, and by making efforts to attract established superior level
statisticians, such as Rudolph Beran from the UC Berkeley and Peter Hall
from Australia. These latter two recruitments became realities soon
after my stepping down as Associate Dean. At the same time, we laid the
foundation for a program in biostatistics, which subsequently developed
into a program of national repute. A decisive role in founding and
developing the biostatistics program was played by Hans Mueller, who was
by training a statistician, a biostatistician and a medical doctor.
Within a few years, it became apparent, and certifiably so, that our
objectives and goals were well on their way of being realized.

I'm sure you recall that, in an evaluation study of 300 statistical
research institutions around the world---carried out by the National
Sciences and Engineering Council of Canada (NSERC) for the period
1986--1990---statistics in the UC Davis was ranked 14th (top~4.7\%)
worldwide, and 11th within the United States (top 3\%). This ranking was
reaffirmed and even improved in a follow-up study, carried out by
Christian Genest and Mireille Guay (\textit{The Canadian Journal of Statistics},
Vol. 30, No. 2, 2002, pages 392--442). In this study, the authors
employed several criteria of evaluating the same as above institutions.
On the basis of one of these criteria---essentially, published research
papers per capita in the ``top 25'' research journals in the field---statistics
at the UC Davis was ranked 4th (top 2\%) among 202
institutions studied. And these hard facts are beyond and above the
general reputation of UC Davis statistics faculty as excellent
researchers, teachers and contributors to the profession. This really is
an achievement for which our faculty as a whole deserves credit, and I
am extremely proud of the hard-working yet congenial group that
constitute the statistics faculty at Davis. Naturally, I am proud as
well of the role I had the opportunity to play in helping to shape this
group.

\renewcommand{\thefigure}{14.B}
\begin{figure*}

\includegraphics{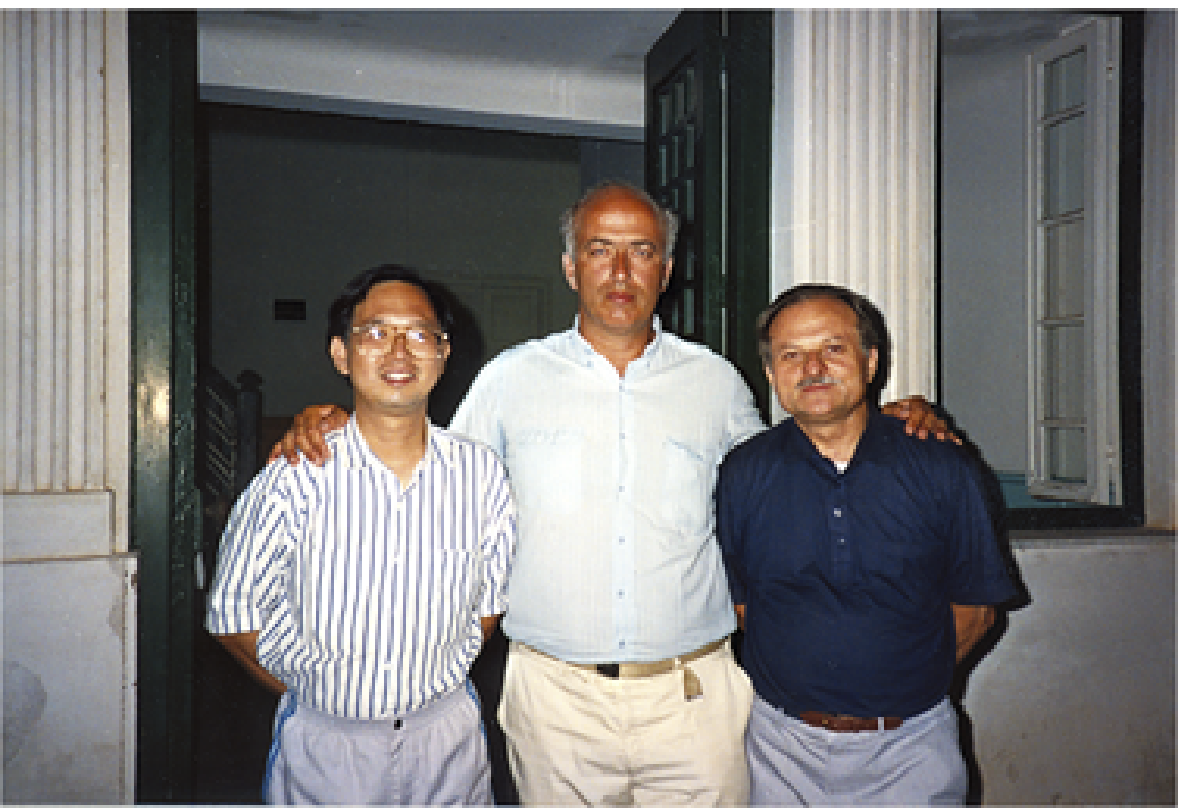}

\caption{In the island of Spetses, Greece, during a NATO Advanced Study
Institute in 1990. From left to right: Y.~P. Mack, UC-Davis; Paul
Deheuvels, L.S.T.A., Universite' Paris VI, France; and George Roussas.}
\end{figure*}
\setcounter{figure}{14}
\renewcommand{\thefigure}{\arabic{figure}}

\textbf{Frank}: George, would you like to mention at this point some
events and turning points we faced as a statistics unit?

\textbf{George}: I surely would. Heading the statistics unit at Davis
was not without its challenges, especially during periods in which the
California economy was weak. In the early 1990's, for example, during
one of the more severe financial crises of the State (with inevitable
repercussions for UC), the dean in charge of day-to-day oversight of the
Intercollege Division of Statistics recommended the merger of statistics
and mathematics as a cost saving device. Not only had the dean (a fine
humanist, but largely unschooled quantitatively) forgotten that that was
where we had started---separating from mathematics in order to realize
the breadth and potential that statistics rarely can achieve within a
mathematics department---but he had failed to reflect on both the
unit's stature and its many applied contributions (including
consultation across the campus through our Statistical Laboratory,
collaborations with applied scientists on campus and a broad spectrum of
courses taught as a service to students in other majors). The faculty
went into overdrive to come up with ideas and strong arguments against
the proposed merger. Also, for a period of about three months, I lobbied
heavily a number of higher-level administrators and other influential
people who were supportive of our continued independence. As a result of
our collective efforts, I submitted a detailed and impassioned letter in
defense of our status as a free-standing unit. In the end, the
administration conceded that the merger of statistics and mathematics
would be a serious strategic mistake. We held our status as an
Intercollege unit for 21 years. While we cherished our considerable
independence as a mini-college on campus, as well as the access it gave
us to various forms of support from all other schools and colleges, we
also realized that we were not big enough to withstand ill-conceived
attacks. It was this kind of reasoning that led us to seek and achieve
the (lesser but safer and, let's face it, more traditional) status of a
department. This went into effect in 2000, and Jane-Ling Wang was the
first chair of the department of statistics. She was succeeded by
Rudolph Beran, and then by the current chair Wolfgang Polonik.

\begin{figure*}

\includegraphics{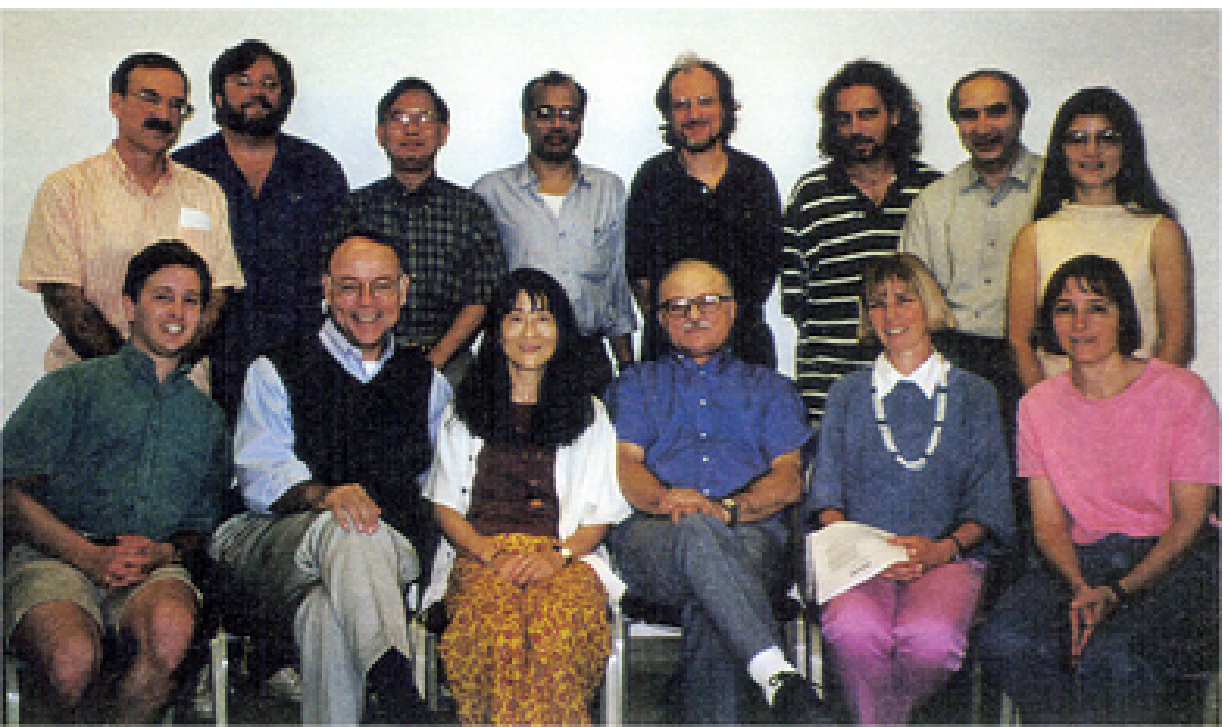}

\caption{Faculty of the Department of Statistics at the UC Davis, Fall
2000. Top line: Alan Fenech, Wesley O. Johnson, Y.~P. (Ed) Mack, Prabir
Burman, Hans-Georg Mueller, Wolfgang Polonik, Rahman Azari and Juanjuan
Fan. Second line: Richard Levin, F.~J. Samaniego, J.-L. Wang, George
Roussas, Christiana Drake and Jessica Utts.}
\end{figure*}

\textbf{Debasis and Frank}: Your friends and colleagues clearly
appreciated your accomplishments and your generous service to several
institutions and to the statistics profession generally. They threw
quite a ``party'' in your honor!

\textbf{George}: All this was something of a surprise to me. My old
friend and former collaborator Madan Puri of Indiana University
organized a volume of research papers featuring work in the general
areas in which my own research was focused. This project resulted in the
Festschrift ``Asymptotics in Statistics and Probability: Papers in Honor
of George Gregory Roussas,'' VSP International Science Publishers, 2000.
It was edited by Professor Puri, and consists of 25 papers by 48 authors
from 13 countries, with a preface co-authored by Madan L. Puri and my
good friend and well-known statistician Yannis Yatracos.

Subsequently, Hans Mueller and the department of statistics conceived of
the idea of organizing a two-day workshop at UC Davis at which the
Fest\-schrift would be officially presented to me. The conference took
place at UC Davis on May 19--20, 2001, with the participation of a select
group of researchers, including a member of the French Academy of
Sciences, a chancellor of a German university, the holder of a
name-chair in the London School of Economics and three chairs of
statistics departments.

\textbf{Frank}: At the time, I made note of the fact that the conference
was scheduled right in the middle of the period that I was on sabbatical
leave in Ireland. I chose not to take offense. (Laughs.) But seriously,
George, I would have loved to have been on hand to give my hearty
congratulations for a roundly successful and meaningful career. And, of
course, the beat goes on.

\textbf{George}: It was indeed unfortunate, Frank, that you could not be
with us on that occasion, but apparently there were many constraints the
organizers had to abide by. I always regretted your absence from that
festive event, in contrast to your ever present continuous support and
counsel throughout my UC Davis years.

\textbf{Debasis}: Let me add, George and Frank, that I was fortunate
enough to be present, and the proceedings were thoroughly enjoyable!

\textbf{Debasis and Frank}: How have you structured your professional
life since leaving the administrative posts you held up to 1999?

\textbf{George}: Well, I continue to be an active member of the
department of statistics, and a member of the graduate program in
statistics and the graduate group in biostatistics, both housed in the
department of statistics. I have concentrated both on a variety of
research problems in the areas in which I've always been interested, and
have enjoyed my teaching assignments, but I have also had the luxury of
time to work on pet projects. I have written three books
[``\textit{An Introduction to Probability and Statistical Inference}'' (2003),
``\textit{An Introduction to Measure-Theoretic Probability}'' (2005) and
``\textit{Introduction to Probability}'' (2007), all published by Academic
Press]. I have already revised the \textit{Measure-Theoretic} book, and I am in
the process of revising another two books. Also, I am in the process of
working collaboratively on a new book (tentative title ``\textit{Probability and
Statistics for Non-majors}''). So, I've kept quite busy, both with
professional projects such as these and with family, to whom we have
added my daughter-in-law Casie, wife to my son John, and my delightful
granddaughter Sophia Aggeliki, as well as my daughter-in-law Laura, wife of my
son Gregory.

\section*{Some Memories from Roussas' Professional Career}

\textbf{Debasis and Frank}: What are your fondest memories over a career
spanning almost 50 years?

\textbf{George}: In retrospect, I feel that there are many reasons that
I should be grateful for my long professional life. I am certainly most
grateful to all of my professors in the UC Berkeley for the truly solid
training imparted in me; the full value of it did not become evident
until later in my academic career. I often reminisce about my ten
productive and pleasant years at the University of Wisconsin, Madison.
Madison was, after all, where I met my wife, Mary Louise, and where our
first son, Gregory, was born. I do not regret the ten to twelve years
that I invested in seeking to contribute to higher education in Greece,
although the net result was almost negligible. I feel that I gave it my
best try, but, realistically, there are many other factors which
influenced the final outcome.

I am certainly most grateful to UC Davis for the way I was received, and
the opportunity I was given to do here what I was not allowed to do in
Greece.

Genuine thanks are also due to my professional colleagues who honored me
with my election as a Member of the ISI (1974), admitted me as a Fellow
in the RSS (1975), and elected me as a Fellow of the IMS (1983) and the
ASA (1986). Special thanks are also due to the scientific community at
large for electing me a Fellow of the AAAS (2008). And last but not
least, I am grateful to a select group of Greek scholars---the membership
of the Academy of Athens---for electing me a Corresponding Member of the
Academy of Athens in the field of Mathematical Statistics (April 17,
2008).

\textbf{Debasis and Frank}: And on a personal level?

\textbf{George}: I feel exceptionally fortunate that I have spent my
adult life surrounded by a wonderful, supportive and endlessly
interesting family. For my sta\-mina and persistence, I must thank Mary,
especially, both for her support and encouragement over the years but
also for her sage advice. I feel singularly fortunate to have three
healthy, intelligent, beautiful sons---Gregory, born in Madison in 1973,
now a computer scientist, John, born in Bloomington in 1977, a
practicing attorney, and George-Alexander, born in Patras in 1980, a UC
Davis graduate in political science, aspiring to the legal profession.
Also, we are delighted with the relatively new arrival (March 26, 2008)
of our first grandchild, Sophia Aggeliki, daughter of John and Casie,
also a practicing attorney and our newest daughter-in-law Laura. Many
wholeheartedly felt thanks are due to
my sisters for their immense moral support and consequential material
support when that was most
needed.

One thing we regret is that we did not have enough time to enjoy the
house that we built in Patras in 1980. Its setting is truly idyllic: It
lies on an acre of land full of trees, (including an olive tree grove)
at the foot of a wooded hill with a mountain in the background, and
faces the Patras bay.

On a personal level, it has also been painful that, by expatriating
myself for most of my adult life, I was deprived of the opportunity to
spend any significant amount of time with my parents, sisters and other
members of the immediate family. In retrospect, I also believe that, by
devoting unduly much time to my professional duties, I deprived my own
family and myself of the opportunity of spending more time together. But
I truly believe that each phase of our lives is a nonrecurrent event,
and must be appreciated, as it comes, to the greatest extent possible.

\begin{figure*}

\includegraphics{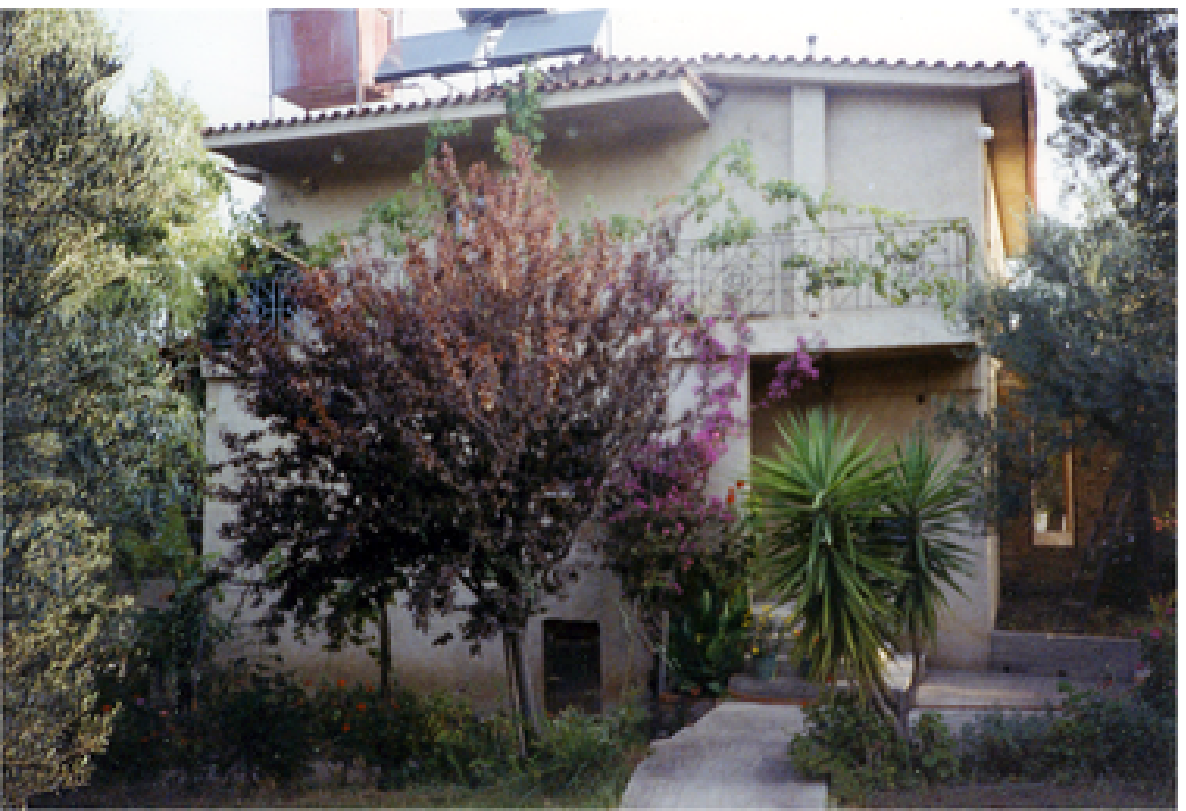}

\caption{One aspect of the Roussas' house close to the university campus
in Patras.}
\end{figure*}

\begin{figure*}[b]

\includegraphics{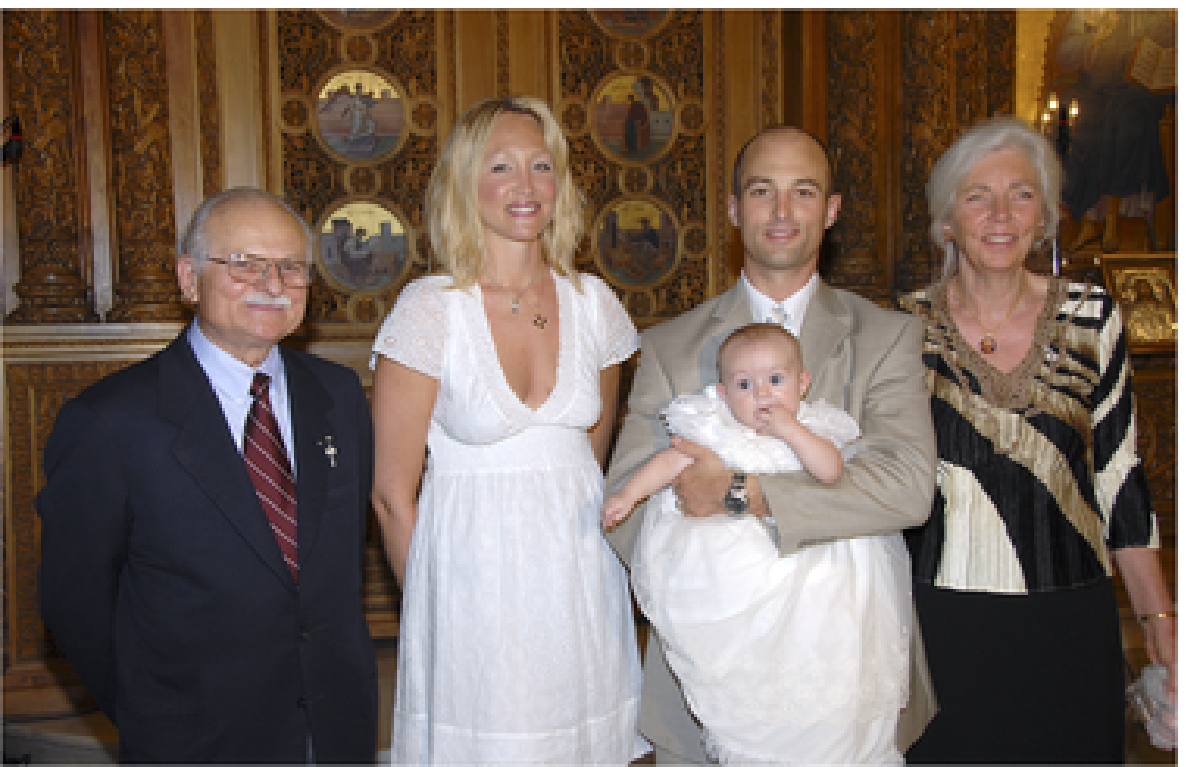}

\caption{At the baptism of the Roussas' first grandchild, Sophia
Aggeliki, in Athens, September 2008. The parents John and Casie Roussas
with the baby, and George and Mary Roussas.}
\end{figure*}

\begin{figure*}

\includegraphics{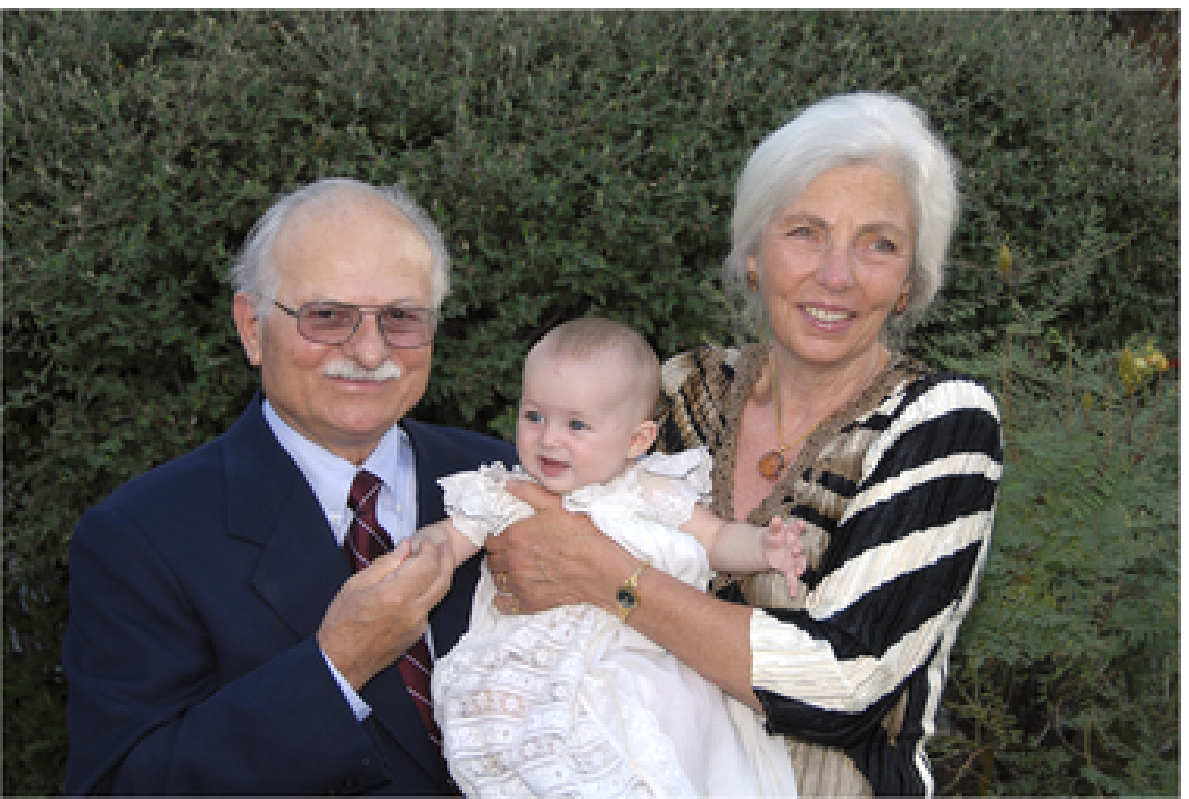}

\caption{George and Mary Roussas with Sophia Aggeliki, right after her
baptism.}
\vspace*{-2pt}
\end{figure*}

\textbf{Debasis and Frank}: Since you've served in a wide variety of
administrative capacities during your acade\-mic career, perhaps you'd
like to share your thoughts about what it takes to do this type of work
well.

\textbf{George}: I am pleased to do so. I believe that being a
good, efficient and inspiring administrator requires an inborn talent.
Beyond this, one has got to be honest, just and straightforward, and by
word and deed, convince one's co-workers about that. One's commitment to
these values must of course be real, but it is also important that they
be clearly perceived by those with whom you deal. Furthermore, without
in any remote way being dictatorial, one has got to convey the clear
message that there is only one leader at a time. One should stand and
defend well established principles, and not bend according to the
prevailing winds. Vision is important, but it is also essential to have
the ability to explain one's vision in ways that gain the needed support
and engagement from others. The ability to listen is extremely
important. It is essential that different sides of a controversial issue
be weighed. I have never found it difficult to take a position that may
be unpopular, but I never wished to do so without being convinced that I
had the relevant facts in hand.

I recall when once working at my desk in the chancellor's office in the
UP, I heard outside my door a rather heated discussion. Inquiring about
it, I was told that it was a committee of cleaning ladies who wanted to
see me and present to me a perennial unsolved issue of theirs, but the
receptionist would not allow them to do so. During my entire tenure on
the university campus, the rumor spread widely about this professor from
the States interacting with people at all levels. However, for the
receptionist it was inconceivable that a cleaning lady would ask to see
the chancellor. On this particular occasion, I invited the committee
into my office, listened to their concerns and was able to resolve them
to their satisfaction that very day.

There were several incidents with highly politicized student and TA
groups, which could have developed to the point of explosion, but
fortunately, they were decisively contained to the point of dissipation.
It is probably best that I not elaborate on them further.

\section*{Brief Description of Main Research Interests}

\textbf{Debasis and Frank}: We haven't spent much time on the areas of
Statistics and Probability that you have concentrated on during your
career. This conversation would be quite incomplete without your giving
us a brief tour.

\textbf{George}: Thanks for asking! As you know---and especially, you,
Debasis---my early work is based on Le Cam's concept of
contiguity and Local Asymptotic Normality (LAN). Roughly speaking, LAN
allows for a more or less arbitrary parametric family of probability
measures to be replaced (asymptotically) in the neighborhood of each
parameter point by an exponential family of probability measures.
Contiguity ensures the establishment of asymptotic normality under
moving parameter points, when such normality under a fixed parameter
point is already known. This theory has important statistical
implications. Roughly speaking, whatever can be done for exponential
families can also be done, in the limit, for the given family of
probability measures. Those results may then be transferred to the
original family, for which they are going to hold at the asymptotic
level. Such results were developed, originally, for discrete
time-parameter Markov processes.

Nonparametric estimation in special cases of Mar\-kov chains has been
around for a long time. However, nonparametric estimation in a general
setting of discrete time-parameter Markov processes was largely an open
area for investigation in the late 1960's. It was exciting to be in on
the ground floor in this problem area. I~published a series of papers,
beginning in 1969, which established some foundational results and
opened the door to further research in the area.

In Markov processes, the future depends on the past and present only
through the present. One way of incorporating the entire past, when that
matters is by
introducing various modes of dependence conditions, referred to as
mixing. The basic idea in mixing processes is that the past and the
future are approximately independent, if they are sufficiently far
apart. In a way, it is the natural evolution beyond Markovian
dependence. I was introduced into this area by my colleague Y.~P. Mack (a
student of Murray Rosenblatt) in 1984, the first year of my visit in the
UC Davis. From a probabilistic viewpoint, there was a huge amount of
work already done, mostly by the Russian probability school (Davydov,
Gorodetski, Ibrahimov, Kolmogorov, Lifshits, Rosanov, Vol\-konskii and
others), and also by probabilists in the States (first and foremost
Rosenblatt, then Bradley, Kesten, Peligrad, Philipp and others), as well
as by other researchers (e.g., F\"{o}ldes, Iosifescu, O'Brien, Withers,
Yokoyama and Yoshihara). However, there wasn't a body of work on
statistical inference on such processes. These problems intrigued me,
and I got some nice results. These were published in a series of papers,
starting in 1987. Ever since, there has been an explosion of papers in
this area, including contributions by Doukhan, Louhichi, Masry, Shao,
Tran, Yu and many others.

The next area of my interest has been that of associated processes. The
concept of associated random variables was introduced by Esary, Proschan
and Walkup in a seminal paper, and it was extensively used in the book
by Barlow and Proschan in a reliability framework. The concept of
negative association was introduced by Joag-Dev. Association was also
introduced and used in the context of mathematical physics by Fortuin,
Kasteleyn and Ginibre.

Although I had a peripheral interest in this area due to my overall
interest in modes of dependence, my interest was accentuated
significantly after an extended visit to UC Davis by Frank Proschan.
Again, there did not seem to have been a systematic approach to
statistical inferences in such processes, and this fact stimulated my
interest in such a kind of work. As a result, there has been a stream of
papers between 1997 and 2001 by me, my students and other collaborators
in which a variety of such problems have been posed and solved. More
importantly perhaps, this seems to have instigated the formation of a
``school'' in this area with much activity in China, South Korea, France
and Portugal. Some of the noted contributors in association, either in
probabilistic developments or statistical inference, have been Birkel,
Bulinski, Cai, Doukhan, Ioannides, Louhichi, Oliveira, Prakasa Rao,
Shashkin, Taylor, Yoshihara and others. Special mention is deserved for
a seminal paper on this subject by C.~M. Newman.

In the last ten years or so, I revisited, with Debasis, the area of
contiguity and LAN, and extended previous work to the so-called Locally
Asymptotically Mixed Normal (LAMN) families of probability measures, so
coined by Jeganathan in 1982. In this latter framework, we produced a
number of papers on distribution theory with applications to statistical
inference.

Finally, my current interests include conditioning, sampling from
continuous time-parameter stochastic processes, and the theory and
applications of copulas.

\textbf{Debasis}: George, I've truly enjoyed the opportunity to work
with you. We've worked on a wide variety of topics, including, of
course, contiguity. Your 1972 book on contiguity has become a classic! I
know it's been translated into Russian and perhaps other languages. Have
you given any thought to writing a new edition of the book that would
include the many new results that we and others have obtained in the
area?

\textbf{George}: The contiguity book, which was published by Cambridge
University Press in 1972, was written in an attempt to obtain a deeper
understanding of the concept of contiguity and its statistical
applications, and also to help disseminate this very important concept.
Le Cam's original paper in 1960 is not particularly easy to read. Of
course, he employed contiguity in his all-inclusive 1986 book
(``\textit{Asymptotic Methods in Statistical Decision Theory in Statistics},''
Springer-Verlag). A much more accessible discussion of contiguity and
its repercussions are presented in the 2000 monograph (``\textit{Asymptotics in
Statistics: Some Basic Concepts},'' 2nd edition, Springer) by Le
Cam and Yang. I was therefore somewhat surprised that the Cambridge
University Press put out (in 2008) a reprint of my book in a paperback
form; apparently, there still seems to be some continuing interest in
that work.

And now, in order to answer directly your question, Debasis: I don't
really have any plans to do what you suggested. However, should you take
the initiative, I might be persuaded to join in! (Laughs.)

Incidentally, some time in the recent past, I had thought of organizing
some material on associated processes and their statistical
applications. This tentative plan is now aborted with the recent
publication of an excellent monograph on the subject matter (``\textit{Limit
Theorems for Associated Random Fields and Related Systems},'' World
Scientific, 2007) by Bulinski and Shashkin.

\section*{Musical Interests}

\textbf{Debasis and Frank}: We know that you have a great appreciation
for classical music. How did that lifelong interest originate, and which
composers are among your favorites?

\textbf{George}: I developed a strong liking for classical music early
on, during my high school days. I'm not sure what drew me to it, other
than its sheer beauty. No one in my immediate environment was
particularly musically oriented. At the same time, I have also always
liked good folk music, as well as some Greek popular music as
exemplified by the two noted composers Hadjidakis and Theodorakis. Also,
I enjoy selected pieces of popular American music and light jazz.
However, my passion is classical music. In general, I am fond of the
Germanic (German--Austrian) composers. I enjoy everything composed by
Beethoven, and, in particular, his third, fifth, sixth and ninth
symphonies, and his fifth (the emperor's) piano concerto. I love many of
Mozart's compositions with special preference for some of his
symphonies, piano concertos 20, 21 and 22, and his operas The Magic
Flute, The Marriage of Figaro, Idomeneo and Don Giovanni. Above all, I
adore his requiem. I very much like a number of symphonies by Brahms and
by Haydn. Also, I enjoy many of Mendelssohn's compositions and, in
particular, the Scottish and the Italian symphonies. I much enjoy the
eternal Messiah by Handel, and on a lighter side, his water music and
royal fireworks. Somewhat surprisingly, I never developed a true liking
of Wagner's compositions, although I have immense appreciation for them.

I also greatly admire many Russian composers. Among them, Tchaikovski
ranks first followed by others, such as Stravinski, Rimsky-Korsakov,
Mussorgsky, Prokofiev, Rachmaninov, Sostakovich and Borodin.

I much enjoy many compositions of the Italian composer Corelli, some of
Vivaldi's compositions, and the arias of operas by Rossini and Verdi. It
would be an omission to leave out my liking of Filandia, and of
symphonies number 2 and 4 by Sybelius, of some compositions by Chopin,
Liszt's Hungarian rhapsody number 2, and also of a couple of pieces by
Bizet. And of course, everybody enjoys Ravel's bolero!

\textbf{Debasis and Frank}: This is clearly more than a~passing
interest. It seems that it ranks right up there~with Jack Kiefer's
appreciation for mushrooms! (Laughs.)

\textbf{George}: Well, you did not ask me about my food preferences!
What a coincidence! I will never pass up---if I can help it---a Saturday
brunch of a mushroom omelet! I do have a preference for certain kinds of
mushrooms, but in the end, any nonpoisonous mushrooms will do!

\section*{General Outlook---Closing Remarks}

\textbf{Debasis and Frank}: On another topic close to your heart, what
are the main tenets of your political outlook?

\begin{figure*}[b]

\includegraphics{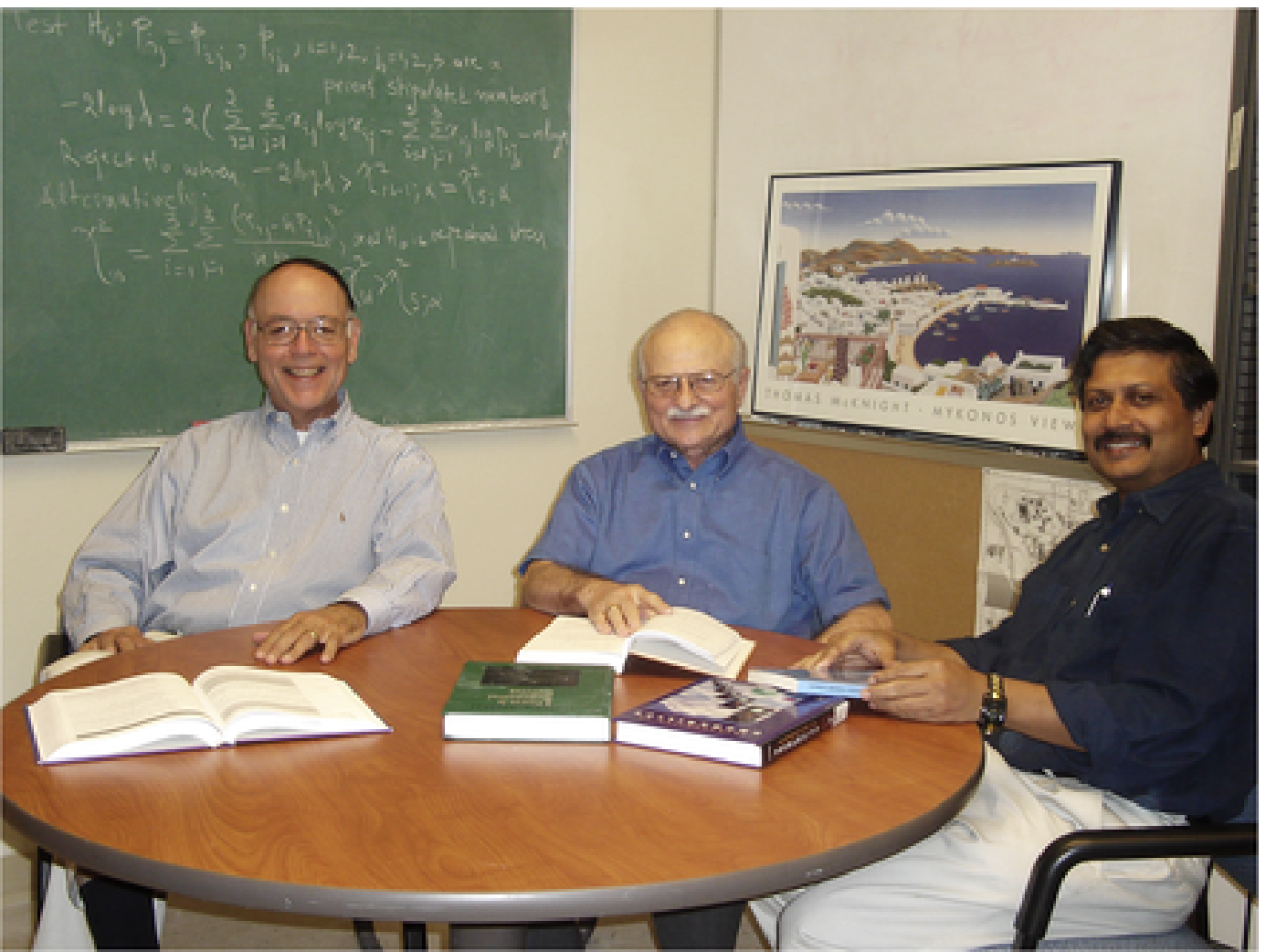}

\caption{Debasis Bhattacharya and Frank Samaniego interviewing
George Roussas (in the middle) in his office in the Department of
Statistics at the University of California, Davis, on May 15, 2009.}
\end{figure*}

\textbf{George}: At least in the recent past and currently, the usual
terms employed to characterize political ideology are those of being
``liberal'' or ``conservative.'' However, these terms are quite
tentative, and have had different connotations in different periods of
time. They are interpreted differently by different people. I like to
think of myself as not really fitting the modern interpretations of
either camp. My basic beliefs are that one should be interested in
preserving the accumulated wisdom of our collective society and in
respecting the greatest intellectual achievements of the human species
over the millennia. If this is this ``conservatism,'' so be it! At the
same time, it seems to me essential that one keeps an open mind and a
positive disposition toward new ideas. That is ``liberalism'' in my
book. However, before a new idea of any importance is adopted, it must
be deeply contemplated and should be vigorously debated and tested. The
novelty of an idea in no way guarantees its worthiness and its
usefulness to society. The unquestioning adoption of ``progressive''
ideas, just because they are novel, may be truly deleterious for the
well being of a society; that is ill-conceived license for perhaps
emotional but certainly not rational behavior, having nothing to do with
liberalism. At the same time, I believe that
it is a mistake to resist
change and adhere to the status quo, simply because it's what we know
and are comfortable with; that is simply reactionary. When I find myself
needing to take a position on a political or social question, I try to
combine what I know or can learn about the alternatives under
consideration and form my opinion based on both experience and newly
found information. In short, I believe
that we should respect existing
social structures, but we should not do so
unquestionably. I have never
been a ``party-line'' type of citizen, and I tend to vote for those
candidates and propositions that seem to me to stand the best chance of
solving real problems and of generally enhancing the quality of our
lives and of the times we live in.

\textbf{Debasis and Frank}: This topic seems to naturally segue into
your general philosophy of life. How would you summarize that?

\textbf{George}: I would say that the most important thing is to live a
``worthy'' life, based on ethical principles, respecting valued
traditions, yet trying to leave things better than you found them.
Trying to make some meaningful contributions to society is important, as
is the avoidance of excesses that distract one from one's more noble
goals and aspirations. I believe that a commitment to ``excellence,'' in
a generalized sense, is also very important. This applies to one's
chosen profession, that is, to the way one does one's work, and also to
one's personal affairs. In both of these areas, integrity and respect
for others, and for society's needs, should always be prime
considerations. Mary and I have striven to raise our sons to have a
sincere appreciation for these same principles.

\textbf{Debasis and Frank}: What advice would you give to young people
just beginning their careers as academicians?

\textbf{George}: Aim high and then work hard, with energy, imagination
and persistence, to achieve your goals. Strive to live a worthy life.
Determine what your main strengths are, and use them to try to make a
difference, both in your professional activities and in your personal
life. Take pride in your best achievements! At the same time, accept
responsibility for whatever failures you encounter and, most
importantly, learn from them.

\textbf{Debasis and Frank}: George, this conversation has been a
distinct pleasure.
You've had an extraordinary career,
with consistently
strong contributions through your research,
your teaching, the highly
respected books and monographs you have written and your many
achievements in administrative capacities. It's been most interesting to
hear about your personal trajectory. You didn't set out with this
trajectory in mind, but we feel very fortunate that it led you to Davis.
We have all benefited from the leadership and collegiality that has
characterized your 25 years here. Thanks for taking the time to talk
with us about your life and career.

\textbf{George}: The pleasure has been all mine. I'd like to express my
deep appreciation to both of you, Frank and Debasis, for this precious
opportunity. Years from now, perhaps my children's children will read
this and be surprised that ``old pappou'' had a~pretty interesting life,
and one that was blessed in~many~ways.

\section*{Acknowledgment}

We would like to thank Patricia Aguilera and
Gloria Anaya for assisting with the preparation of the manuscript.

\end{document}